\documentstyle[psbox]{WORKSHOP}
\begin{document}

% TITLE OF THE PAPER
%  If the title is too long for a single line, you can split it 
%  by putting two backslashes. 
%  You might want to put the subtitle. Then it should be inserted 
%  within {\large\sf  }.
%  e.g.:  
%     \title{ Too Long Title \\ for one line \\
%     {\large\sf Subtitle} }
\title{
MAXI and GLAST studies of Jets in Active Galaxies \\ 
%{\large\sf  -- Brief Instructions for Users of the `WORKSHOP' Style File --} % SUBTITLE
}

% AUTHOR(S) 
\author{
Greg Madejski$^1$, Jun Kataoka$^2$, and Marek Sikora$^3$\\ 
\\[12pt]  % TO BE SPACED WITH ONE LINE
%
% INSTITUTES OF AUTHORS
$^1$  KIPAC and SLAC, Stanford University, Stanford, CA 94305 \\
$^2$  Tokyo Institute of Technology, 2-12-1, Ohokayama, Tokyo, Japan \\
$^3$ Copernicus Astronomical Center, Warsaw, Poland \\
%$^3$  Institute and its address of author4 and author5 \\
%
% please put the first author's initial and e-mail address below
{\it E-mail(GM): madejski@slac.stanford.edu} 
%            \_ Initial      \
%                             \_ E-mail address
}

\abst{
The recent launch of GLAST - coinciding with the MAXI workshop - opens a new
era for studies of jet-dominated active galaxies, known as blazars.  
While the emission processes operating in various spectral bands in blazars 
are reasonably well understood, the knowledge of the details of the 
structure of the jet, location of the dissipation region with 
respect to the accreting black hole, and coupling of
the jet to the accretion process are known only at a rudimentary level.  
Blazars are variable, and this provides an opportunity to use the variability 
in various bands - and in particular, the relationship of respective 
time series to each other - to explore the relative location of 
regions responsible for emission in the respective bands.  Observationally, 
this requires well-sampled time series in as many spectral bands as possible.  
To this end, with its all-sky, sensitive monitoring capability, the recently 
launched GLAST, and 
MAXI, to be deployed in 2009, are the most promising instruments 
bound to provide good sampling in 
respectively the energetic gamma-ray, and the soft X-ray band.  
This paper highlights the inferences regarding
blazar jets that can be gleaned from such joint observations.  
}

\kword{galaxies: active -- X-rays: galaxies -- galaxies: quasars}

\maketitle
\thispagestyle{empty}

\section{Introduction:  MAXI and active galaxies}
The MAXI All-sky monitor, planned to be deployed on the International 
Space Station in 2009, will be the 
preeminent facility to monitor the entire sky in the soft X-ray band.  
The most important extragalactic targets for MAXI will be active galactic 
nuclei, often strongly variable in the X-ray band.  
There, the well-sampled time series for many objects - and 
the resulting Power Density Spectra - 
are bound to significantly advance our understanding of the central engines of 
AGN.  In particular, those are likely to provide 
an independent estimate of the masses and mass 
function of the central black holes, 
which can be otherwise difficult to obtain via other means, especially 
for objects at a considerable redshift.  

Besides measuring the properties of the time series themselves, 
the MAXI data will be tremendously useful for cross-correlation 
of the well-sampled time series against those measured in other 
bands.  One class of AGN where such measurements will be particularly 
valuable are those active galaxies where the dominant portion of the 
observed radiation is produced in a relativistic jet pointing close to our 
line of sight.  Such objects are known as blazars, and those are the 
objects where the MAXI - GLAST\footnote{At the time of
preparation of this manuscript, GLAST - or Gamma-ray Large Area Telescope - was renamed to be the Fermi Gamma-ray Space Telescope.  Since the conference took place prior to renaming, for consistency, this paper will use the name ``GLAST'' throughout.}  synergy will be most apparent.   They generally 
are characterized by broad-band spectra often extending to the 
highest observable regimes such as the GeV or even TeV band, 
coupled with large amplitude, chaotic variability seen in all bands;  
strong polarization in the radio, optical, and IR;  and the presence of strong 
radio emission arising from extremely compact ($\sim$ milliarcsec) and often 
physically variable structures imaged with the Very Long Baseline 
Interferometry (VLBI), often associated with the so-called apparent 
superluminal expansion.  It is in fact the flux measured in the GeV or TeV 
band that often dominates the overall energetics.  Clearly, time-resolved 
gamma-ray observations, combined with monitoring in other bands, 
are important for detailed studies of blazars, with the goal of understanding 
the emission mechanisms, leading in turn to determination of the content of 
radiating particles, then the structure of the jet (the energy dissipation 
mechanism) and ultimately its connection to the central engine, 
presumably powering the blazar phenomenon via accretion onto the 
supermassive black hole.  

\begin{figure*}[t]
\centering
%\psbox[xsize=0.4#1,ysize=0.2#1,rotate=r]
\psbox[xsize=8cm]
{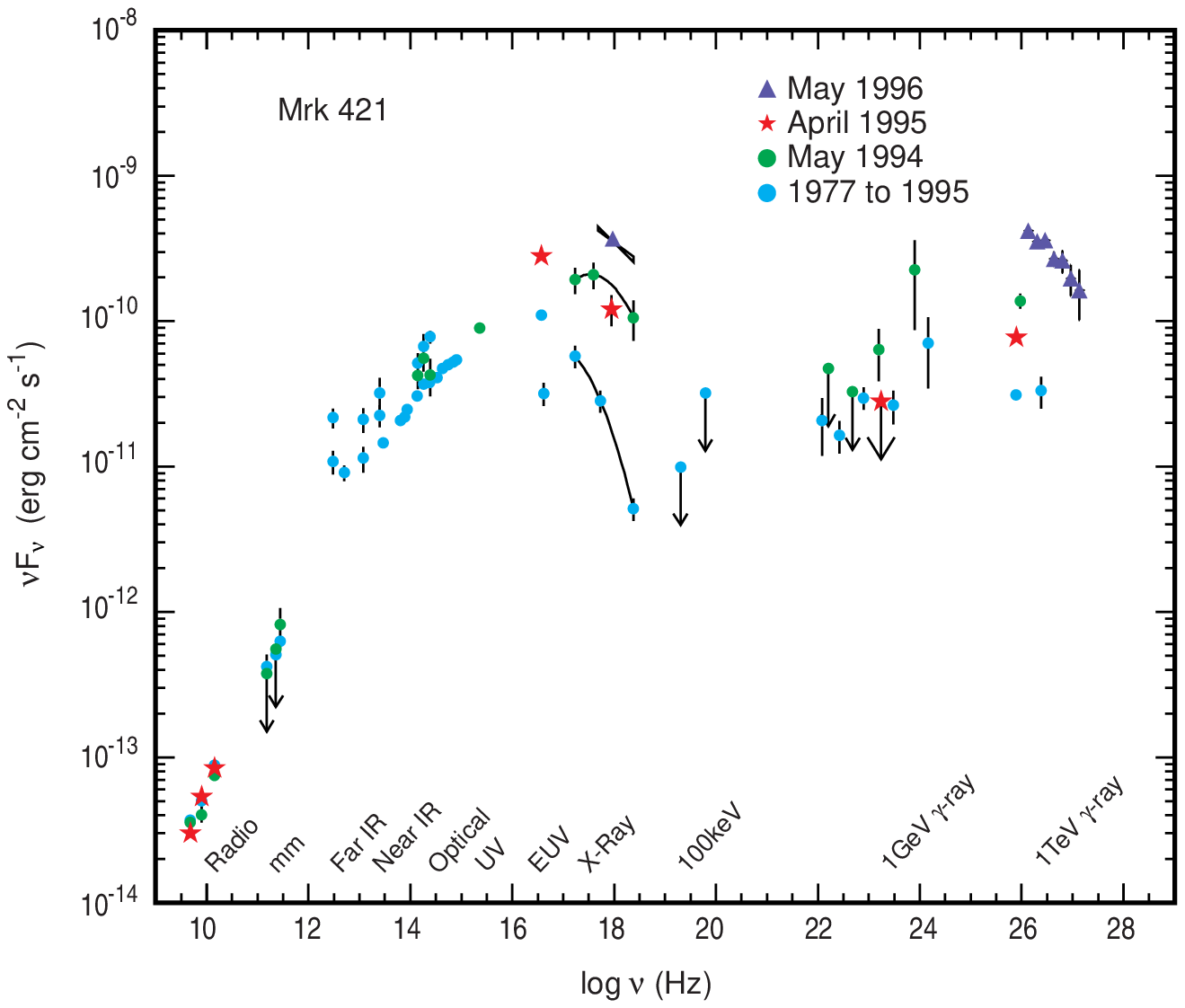}
\hskip 2 cm
\psbox[xsize=7cm,rotate=r]
{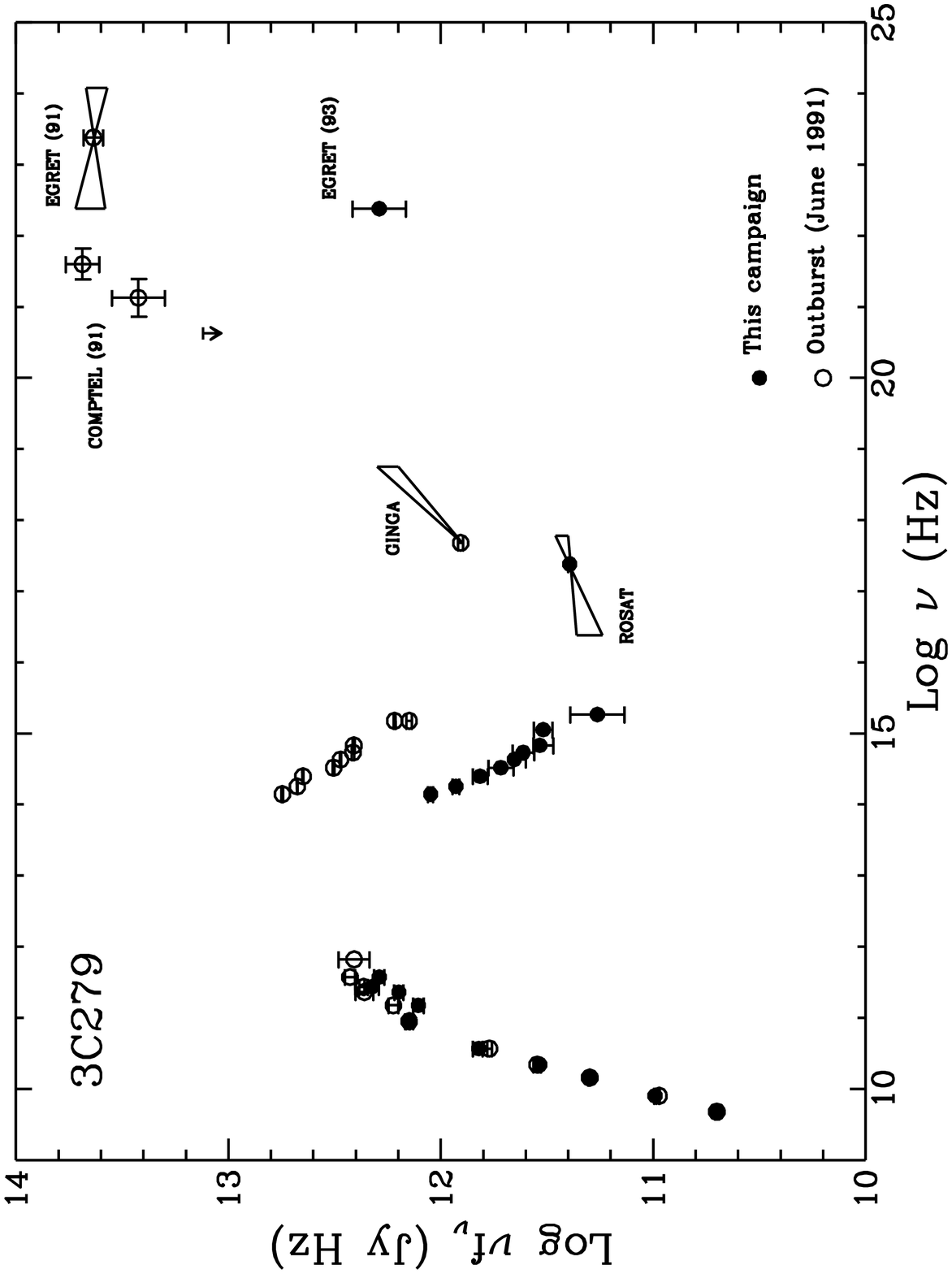}
\caption{ Examples of broad-band spectra of two blazars.  Left panel 
illustrates the broad band spectrum and variability range in various bands of 
Mkn 421, the first blazar detected unambiguously in the TeV band
(Macomb et al. 1995).  Right panel illustrates broad-band spectrum of 3C279, 
a blazar associated with quasar at $z=0.524$, detected as a bright 
gamma-ray source with EGRET (Kniffen et al. 1993).  The data are from 
separate multi-band campaigns, reported in Maraschi et al. (1994) and 
Wehrle et al. (1998)}
\end{figure*}

\section{Gamma-ray Large Area Telescope and MAXI as the key tools for studies of blazars}
GLAST's main instrument, the Large Area Telescope, or LAT, relies 
on the conversion of gamma-rays into electron-positron pairs;  
tracking of those pairs allows the determination of the direction 
of the incident gamma-ray.  Such a design results in 
a wide solid angle of the sky, $\sim 2$ steradians, 
simultaneously available to the detector.  During the normal 
operation, GLAST points away from the Earth, and slight 
rocking of the axis of the spacecraft 
allows monitoring of the entire sky on time scales shorter than 
a day-long or even shorter, since the whole sky 
is surveyed in $\sim 3$ hours.  GLAST's bandpass covers $\sim 20$ 
MeV to $\sim 300$ GeV, featuring a peak effective area (at $\sim 1$ GeV) 
of $\sim 10,000$ cm$^2$.  The point spread function of the instrument 
(corresponding to 68\% containment) is about $0.5^{\rm o}$ at 1 GeV, 
and the energy resolution is about 10\% (the details of the instrument 
are given in Atwood et al. 2008).  Those parameters are significantly 
better than the LAT's predecessor, the EGRET instrument onboard of the 
Compton Gamma-ray Observatory, allowing studies for many more blazars at 
shorter time scales than previously possible.  

\section{Diversity of blazar emission processes in the X-ray band} 
Why is then the X-ray band important for understanding of blazars?  
Bulk of the emission arises in the Doppler-boosted relativistic jet;  
in a sub-class of blazars showing quasar-like properties, a (generally) 
small fraction of the optical and UV light is also detected to be 
emitted nearly isotropically, in the accretion disk, and often reprocessed in 
the broad emission line region.  The Lorentz factors of the jets 
$\Gamma_{\rm jet}$ are measured via superluminal expansion, and are 
roughly $\sim 10$ (Jorstad et al. 2001).  The broad-band spectra 
(plotted in the log ($E \times F(E)$) vs. log$(E)$  form) of two 
representative objects studied extensively, 3C279, and Mkn 421, 
are shown in Figure 1.  Very broadly, the emission consists of two peaks, 
one with a maximum between the far IR and soft X-ray band, and another 
in the gamma-ray band.  Since the low energy peak often shows considerable 
polarization (measured so far in the radio, IR and optical bands), it is 
generally agreed upon that the dominant emission mechanism responsible 
for radiation in the low-energy peak is the synchrotron process of 
ultra-relativistic electrons accelerated in magnetic field. 
Most modeling suggests that the Lorentz factors of electrons 
$\gamma_{\rm el}$ radiating near the peak are around $10^3 - 10^4$ 
for the (generally higher luminosity) blazars associated with quasars 
(such as 3C279 in Fig. 1), and $10^5 - 10^6$ for line-devoid, generally 
lower luminosity objects (such as Mkn 421 
in Fig. 1), which are often sources of TeV gamma-ray emission.  
The high energy peak, on the other hand, is generally 
attributed to Compton scattering by the same population of electrons 
that produce the low energy peak, with target photons being the ambient, 
external, diffuse (quasi-isotropic) radiation field 
in the former class, and the synchrotron 
radiation internal to the jet in the latter class.  Probably the most
compelling reason for the difference between the two classes is the 
mass accretion rate (in Eddington units).  In the latter class, the 
accretion corresponding to a low $L/L_{\rm Edd}$ forms a ``hot'' 
flow, where the density of the accreting material is never 
sufficiently high to allow efficient cooling and formation of 
a ``standard, '' ``cold'' - and thus luminous 
accretion disk, and therefore the energy is advected to 
the black hole with the falling matter.  In the former class, $L/L_{\rm Edd}$
might be higher, allowing for a ``cold'' (roughly $10,000 - 30,000$ Kelvin), 
luminous disk to form - and thus we detect the 
quasi-isotropic signatures of accretion such as the emission lines, and 
sometimes even the ``blue bump.''

As is apparent from Figure 1, the minimum between the two peaks 
is located in the X-ray band.   In the former sub-class (as in 3C279), 
the optical / UV spectrum, being generally quite steep (soft), is presumably 
the ``tail'' of the synchrotron component, while the X-ray 
emission - generally showing relatively hard spectra, with energy indices 
$\alpha_{\rm X} < 1$ - suggests an association with the onset of 
the Compton component.  There, at least in the context of the models 
considered above, the X-ray emission is due to the {\sl low energy 
end} of the electron energy distribution.  Since the observed X-ray spectra 
are not measured to be harder than $\alpha_{\rm X} < 0$ - this suggests 
that the X-ray band probes the ``most populous'' part of the electron 
population, where the electron number density is the greatest.  
With this, in the blazars associated with 
quasars, the {\sl X-ray spectral measurements are 
crucial in determining the total content of the radiating
particles in the jet}.  As an aside, in principle, such a measurement 
could be performed in the low-frequency radio regime, since the 
low-frequency radio and X-ray bands both mirror the low energy 
electron population.  There, however, the synchrotron self-absorption makes 
the lowest energy portion of the jet inaccessible, and since the inner jet 
is thus optically thick, the radio observations can penetrate only down 
to a ``photosphere'' and thus probe considerably more extended spatial 
region.  

The situation in the latter class of blazars - such as Mkn 421 - 
is the opposite.  There, the low-energy (synchrotron) component peaks in the 
optical/UV or even the soft X-ray regime;  
the X-ray spectra
are generally much softer (with $\alpha_{\rm X} \sim 1$ or softer) 
than in blazars associated with quasars and 
represent the high energy portions/tails of the synchrotron spectral
component.  Here, the X-ray band 
probes the highest energy ``tail'' of the electron distribution.  
Detailed studies in this band are indispensable in determining the 
extent of the energy distribution of the most energetic radiating electrons, 
which in turn is needed to provide the strongest constraints
on the particle acceleration mechanisms.  

\vskip -2 cm

\begin{figure*}[t]
\centering
%\psbox[xsize=0.4#1,ysize=0.2#1,rotate=r]
\psbox[xsize=8cm]
{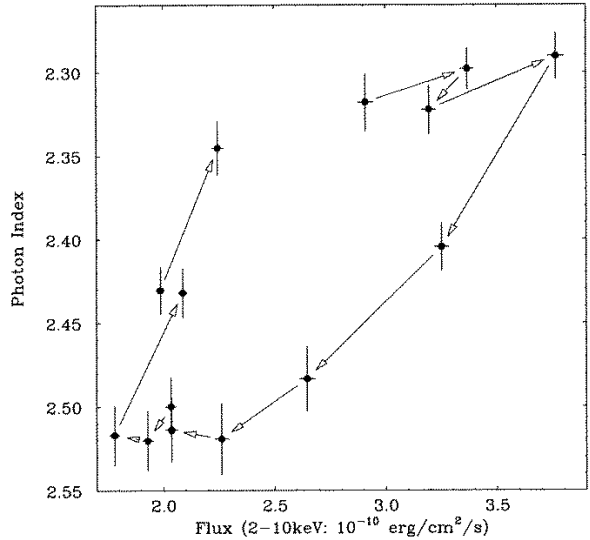}
\hskip 1 cm
\psbox[xsize=6.5cm]
%{Mkn421_hyst.ps}
{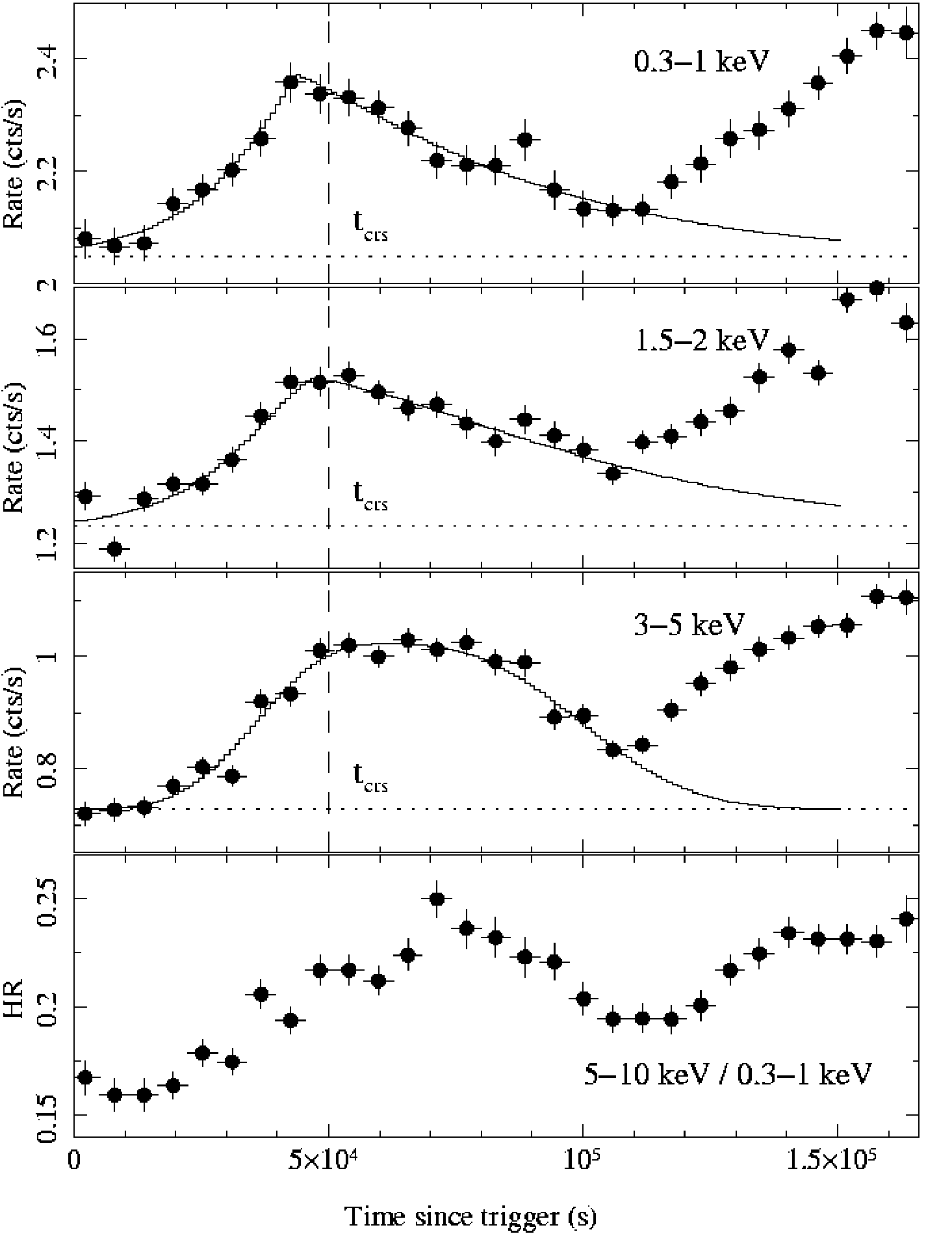}
\caption{ Results of time-resolved X-ray spectroscopy of blazars.  
Left panel shows the evolution of the spectrum of blazar 
Mkn 421 observed in 1994 with the ASCA satellite, with the X-ray 
spectrum becoming harder when the source increased in intensity;  
the total length of the observation was about 1 day (from Takahashi 
et al. 1996).  Right panel shows the Suzaku observation of another 
TeV-emitting blazar, 1ES1218+304.  There, the spectral variability 
had an opposite sense:  the source become brighter in soft X-rays, 
while the hard X-ray flare peaked $\sim 20$ ks later (from Sato et al. 2008).  
}
\end{figure*}

\section{Two facets of time variability}
High energy emission from essentially all active galaxies is variable, 
which provides a difficulty but also an opportunity.  On one hand, 
a reliable measurement of the broad-band spectrum must be obtained 
simultaneously, but on the other hand, the time variability in 
any single band - but also, a relationship of such variability 
amongst various bands - provides an opportunity to explore causal 
relationships between the respective emission processes and/or 
emission regions.  With this, there are clearly 
two separate aspects - and associated challenges - of 
time variability studies of any astronomical sources.  One is the
sensitive measurement of properties of the time series in any single band, 
while the other is cross-band correlations.  

\subsection{Intra-band variability}
The ``standard'' approach for the former aspect is to measure the 
Power Density Spectrum, which is essentially a Fourier transform 
of the time series.  Another is the Structure Function;  both were 
covered in detail in presentations at the Workshop, including those 
by McHardy, Hayashida, and past presentations by Kataoka, Edelson, 
Markowitz and others:  in reference to blazars, see, in particular, 
a review by Kataoka (2008).  Over a narrow range, PDSs or SFs 
of such time series 
are generally well-described by a power law, but the index of the power 
law changes with the variability frequency, at some ``break'' frequency 
$f_{\rm break}$.  In accreting black holes - where the emission is 
presumably mainly due to nearly-isotropic radiation from the accretion 
disk - this break frequency has been demonstrated to correlate well with 
the mass of the accreting object (see, e.g., 
the relevant figure in McHardy, these 
proceedings, an extension of earlier work by Hayashida et al. 1998).  
Remarkably, this relation seems to hold down to masses of stellar size 
black holes.  

Studies of jet-dominated blazars are quite sparse, with 
the first robust report for a sample of X-ray bright BL Lac 
objects by Kataoka et al. (2001), suggesting a 
break on day-long time scales, but with no clear correlation with the
black hole mass (which, in turn, is difficult to determine in blazars).  
So far, reports for a clear PDS break are limited to one object, 3C273 
possessing at least some blazar-like characteristics, but also showing 
substantial contribution from the emission from the accretion disk (McHardy, 
these proceedings;  see also I. McHardy's presentation at the Blazar 
Variability Workshop in Paris, April 2008).  There, 
the break frequency seems to follow that expected from the BH mass.  Clearly, 
more detailed studies of blazars are needed, to determine their
PDSs, compare those to the results for non-jet AGN, and determine whether 
there is any clear correlation with the black hole mass.  Such studies are 
challenging, since they require well-sampled monitoring over a long span of 
time, corresponding to many years.  MAXI is ideally 
positioned for this task.  

\subsection{Inter-band variability}
The inter-band variability studies are equally difficult, again, because of 
the severe effect of sampling on the robustness of determination of any lags 
or leads.  With this, since many instruments provide data for a single 
source at a time, planning, scheduling, and coordination of various 
facilities is quite complex and challenging.  Nonetheless, there are some 
recent successes resulting from well-planned campaigns:  one notable 
example using a large suite of telescopes ranging from radio to X-ray 
bands, and including optical polarimetry, is the variability study of 
BL Lacertae (Marscher et al. 2008).  There, the data indicate that there 
are multiple dissipation regions in the jet, and at least some of the X-ray 
flux arises at a distance $\sim$ thousands of $R_{\rm S}$ from the black 
hole.  This observational result has many implications on the structure 
and the content of the jet, and similar observations need to be conducted 
using other objects.  

Another important observational result concerns monitoring the TeV-emitting 
blazars, such as Mkn 421 mentined in Sec. 3.  There, at least in the context 
of widely accepted leptonic models, the X-ray band and the TeV band both 
should reflect the high-energy end of the radiating electron population, 
so variability patterns should be correlated.  While this is generally 
the case, the amplitudes of flares detected in the respective X-ray and 
TeV bands do not necessarily follow the trend expected in the simplest 
scenarios (for a recent discussion, see, e.g., Fossati et al. 2008).  
Clearly, more detailed understanding of those objects will require 
good temporal coverage in the X-ray band.  

Meeting those goals requires overcoming rather severe observational 
challenges.  The determination of the PDSs and inter-band correlations 
alike is difficult because of strong detrimental effect of the sampling 
pattern on the resulting PDS and cross-correlation function, as discussed 
in the last Section below.  Data are sparse, 
because most X-ray facilities observe one object at a time 
(or, at most one field with a bright blazar at a time).  
With this, MAXI, monitoring all sky simultaneously, provides much better 
sampled X-ray data for a number of objects.  Still, 
it is important to note that MAXI is sufficiently 
sensitive only for a limited number of blazars, since blazars often can be 
relatively faint in the X-ray band (see above).  Still, given the need for 
good sampling, a temporal coverage of MAXI is indispensable.  

\begin{figure*}[t]
\centering
%\psbox[xsize=0.4#1,ysize=0.2#1,rotate=r]
\psbox[xsize=19.0cm]
{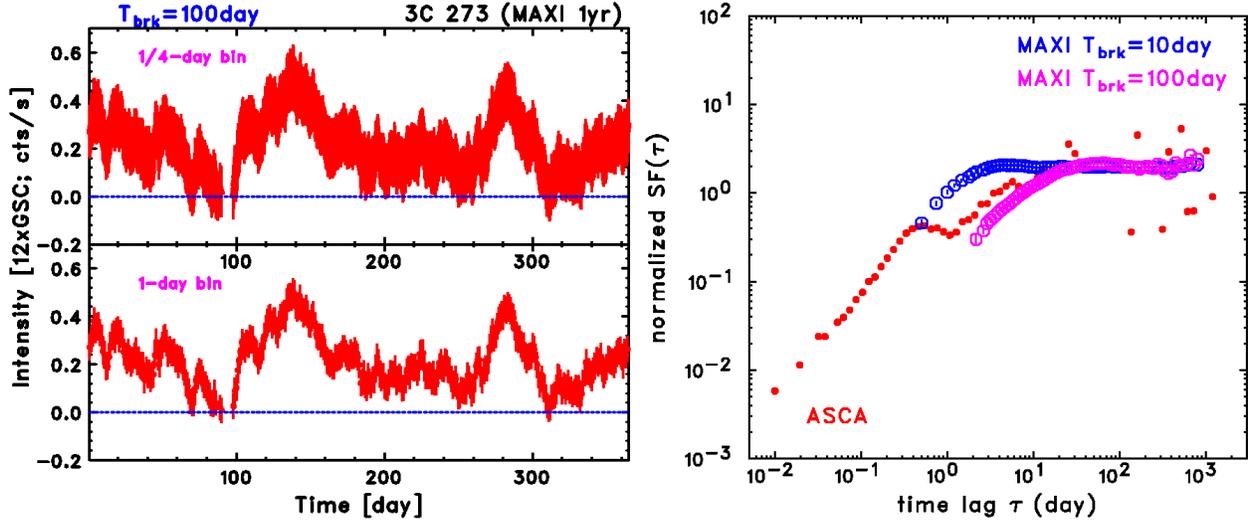}
%\hskip 2 cm
%\psbox[xsize=8cm,rotate=r]
%{3C279_sp.eps}
\caption{ Right panel:  Simulated X-ray light curve of 3C273 
as observed with MAXI.  Data are binned at 1/4 day (upper plot) and 1 day 
(lower plot) intervals.  Right panel:  Structure Function for the blazar 
Mkn 421 calculated from the simulated light curves obtained with MAXI;  
the assumed mean flux is 10 milliCrab.  For comparison, the plot also includes 
SF derived from combining the Asca and RXTE ASM data. (Figure from 
Kataoka 2008)}
\end{figure*}

\subsection{Recent intra-band spectral variability studies of blazars:  Suzaku observation of 1ES 1218+304} 
The cross-band variability studies show great promise as a tool 
to study the details of structure of blazar jets emitting over 
a broad range of spectral bands as above.  But even studies of 
spectrally-resolved variability in a single band (such as, e.g., wide-band 
X-ray observations) can be very fruitful.  Among the most important results 
regarding the nature of the radiating particle distribution in blazars was the 
Asca observation of the X-ray spectral variability in the TeV-emitting blazar 
Mkn 421, with the broad-band spectrum shown in Fig. 1.  There, the X-ray 
spectrum corresponds to the most energetic tail of the electron 
distribution, and the Asca data revealed that the X-ray spectrum 
became harder as the object became brighter, with hardening of the 
spectrum occurring more rapidly than the increase of brightness.  More
importantly, in the decay phase, the spectrum became softer more 
rapidly than the flux decrease (Takahashi et al. 1996;  see Fig. 2, 
left panel).  At least regarding the flux decay phase, 
this is precisely the kind of spectral variability expected 
in the synchrotron or Compton process, where the energy loss is energy 
dependent.  The knowledge of the characteristic synchrotron cooling 
time $\tau_{\rm cool} (E)$ - going as $\tau_{\rm cool} \propto E^{-1/2}$
(derived from the cooling time scale for electrons 
$\tau_{\rm cool}(\gamma_{\rm el}) \propto \gamma_{\rm el}^{-1}$)
- allowed an independent constraint on the Lorentz 
factors of the radiating electrons $\gamma_{\rm el}$, and, in fact, yielded 
values of $\gamma_{\rm el}$ similar to those inferred from broad-band 
spectral fitting.  

Of course the implicit assumption in the analysis above is that the 
acceleration process is very rapid, significantly more rapid than the 
electron energy loss time scale even at the highest observable energy 
(note that the acceleration time scales for more energetic electrons 
are longer, with the dependence of the acceleration time scale
on the photon energy going as $\tau_{\rm acc} (E) \propto E^{1/2}$, derived 
from $\tau_{\rm acc}(\gamma_{\rm el}) \propto \gamma_{\rm el}$).  
This, in turn, might depend on a particular blazar, or even on 
a particular event.  In fact, an opposite behavior to that described 
above for Mkn 421 was detected recently with Suzaku in another TeV-emitting 
blazar 1ES1218+304, via observations reported by Sato et al. (2008).  
There, the $\sim 2$ day long observation revealed a well-resolved 
X-ray flare with a rise time of $\sim 50$ ks.  The spectrally 
resolved Suzaku light curve (see Fig. 2, right panel) indicated that the 
hard X-ray (5 - 10 keV) flux in the flare clearly lagged that in 
the soft X-rays (0.3 - 1 keV) by about 20 ks.  This was associated with 
the profile of the flare, which became more symmetric at higher energies.  
This lag can now be interpreted as an energy-dependent signature of the 
electron {\sl acceleration} process.  Those two features of the 
energy-resolved time series now suggest that the acceleration and 
cooling time scales in this object are roughly comparable for electrons
radiating at $\sim 2$ keV - and this puts further constraints on the 
electron acceleration process, and in particular, on the level 
of turbulence in magnetic field (see Sato et al. 2008).  

\section{Future facilities to study blazars in the X-ray band}

The launch of GLAST is taking place when several X-ray sensitive 
observatories are operational and delivering high quality data, 
but also with several additional ones under construction.  Most notably, the 
Chandra satellite, featuring the bandpass of $\sim 0.5 - 10$ keV, 
is providing probably the most sensitive measurements on many sources.  
Important discovery with Chandra - mainly owing to the superb quality, sub-arc 
sec imaging - was the detection and mapping of the kpc size X-ray 
emitting jets in a number of blazars, including those that 
are bright gamma-ray emitters detected with EGRET.  XMM-Newton observatory 
has capabilities in many ways similar to Chandra's, with somewhat worse 
point spread function of its three mirrors, but with substantially 
better effective area, allowing sensitive measurements of flux variability of 
active galaxies on even shorter time scales than Chandra.  
Just as is the case for Chandra, XMM-Newton is also in a deep 
(several day-long) orbit, allowing uninterrupted data streams and unambiguous 
variability studies on time scales of a $\sim$ day down to the Poisson limit, 
generally corresponding to a few hundred seconds.   
Suzaku has possibly the best effective area for its low-energy (XIS) 
detectors, and also features an additional, hard X-ray detector 
- extending the bandpass (for most AGN) to at least 50 keV;  
however, the low-Earth orbit and Sun angle constraints place some 
limitations on the sampling pattern and the resulting measurements 
of the intra- and cross-band variability properties.  Swift satellite 
features nearly all-sky pointing capability, and is bound to be 
an important tool for rapidly responding to, and following exceptional 
flares of blazars.  In the near future, NuSTAR, sensitive in the 
hard X-ray ($\sim 10 - 80$ keV) band, slated to be launched in 2011 or 2012, 
will be monitoring the hard X-ray flux for at least a 
few blazars for $\sim$ weeks, but both Swift and NuSTAR, with their 
low-Earth orbits, will be subjects to periodic source occultations. 
All those facilities, however, can observe only {\sl one target at a time}.  
With the Rossi X-ray Time Explorer - with its All-Sky Monitor - 
nearing the end of its operational life, a working all-sky monitor 
is essential, and that capability will be provided by MAXI.  This 
will be important for GLAST, but equally, or even more so, 
for the current and future TeV-sensitive Cerenkov telescopes.  
MAXI will provide a trigger and monitoring capability for observations 
of the TeV-emitting blazars regardless whether they are or are not 
subjects on on-going multi-band campaigns.  

What are MAXI's capabilities to measure variability of blazars?  
Bright objects such as 3C273 ($\simeq 5$ milliCrab) and Mkn 421 
($\simeq 10$ milliCrab) can be detected with more than 5 $\sigma$ 
level every day, allowing for the first time non-bias monitoring 
of the sources from daily to more than yearly time scales.  
Figure 3 (left panel) shows the simulated long-term (1 year) 
light curve for 3C273, assuming a PSD slope of 2.0 with a break time of 
$1/f_{\rm break} \simeq 100$ day.  The resultant structure function 
(calculated as in Kataoka 2008) is illustrated in the right panel 
of Figure 3:  it clearly reveals variability nature of blazars on long 
time scales, significantly longer than the characteristic break.  
Detailed measurements of such features in the PDS and SF are bound to 
probe whether the variability properties are mainly governed by the 
changes in the accretion flow, or by the properties of the 
energy dissipation
process and attendant instabilities in the relativistic jet.  

\vskip 0.7 cm

\noindent {\bf Acknowledgements:} The lead author is grateful to the 
conference organizers for 
their hospitality and support, and acknowledges valuable comments from Drs. 
Benoit Lott and Luigi Costamante, as well as support via the 
Department of Energy Contract
DE-AC02-76SF00515 to the Stanford Linear Accelerator Center.

\section*{References}

\re
Atwood, W., et al. 2008, submitted to ApJ

\re
Fossati, G., et al. 2008, ApJ, 677, 906

\re 
Hayashida, K., et al. 1998, ApJ, 500, 642

\re
Jorstad, S., et al. 2001, ApJS 134, 181

\re
Kataoka, J. et al. 2001, ApJ, 560, 659

\re
Kataoka, J. 2008, in Proceedings of Science, Proc. 
of Workshop on Blazar Variability Across the Electromagnetic 
Spectrum, Ed. B. Giebels, arXiv:0806.4243
 
\re
Kniffen, D. et al.  1993, ApJ 411, 133

\re 
Macomb, D., et al. 1995, ApJ, 459, L111

\re
Maraschi, L., et al. 1994, ApJ, 435, L91

\re
Marscher, A. P., et al. 2008, Nature, 453, 966

\re
Sato, R., et al. 2008, ApJ, 680, 9L

\re
Wehrle, A., et al. 1998, ApJ, 497, 178

\label{last}

\end{document}